\documentclass[conference]{IEEEtran}

\usepackage[hyphens]{url}
\usepackage[sort]{cite}
\usepackage[usenames, dvipsnames]{color}
\usepackage[nolist,nohyperlinks]{acronym}
\usepackage[table,xcdraw]{xcolor}
\usepackage{tabularx}
\usepackage[pdftex]{graphicx}
\DeclareGraphicsExtensions{.pdf,.jpeg,.png}
\usepackage{array}
\usepackage{caption}
\usepackage{subcaption}
\usepackage{multirow}
\usepackage{algorithm}  
\usepackage{algorithmicx}  
\usepackage{algpseudocode}  
\usepackage{amsmath}

\usepackage[
top    = 1.70cm,
bottom = 1.05in,
left   = 0.63 in,
right  = 0.63 in]{geometry}

\IEEEoverridecommandlockouts

\newcommand{\PreserveBackslash}[1]{\let\temp=\\#1\let\\=\temp}
\newcolumntype{M}[1]{>{\PreserveBackslash\centering}m{#1}}
\newcolumntype{C}[1]{>{\PreserveBackslash\centering}p{#1}}
\newcolumntype{R}[1]{>{\PreserveBackslash\raggedleft}p{#1}}
\newcolumntype{L}[1]{>{\PreserveBackslash\raggedright}p{#1}}

\newcommand{\fakepar}[1]{\smallbreak\noindent}

\usepackage{manyfoot}
\DeclareNewFootnote{A}
\DeclareNewFootnote{B}

\let\footnoteR\footnoteB
\let\footnote\footnoteA
\iffalse
    \newcommand{\mb}[1]{\footnoteR{{\color{red}\bf MB: #1}\color{red}}}
    \newcommand{\yj}[1]{\footnoteR{{\color{red}\bf YJ: #1}\color{red}}}
\else
    \newcommand{\mb}[1]{}
    \newcommand{\yj}[1]{}
\fi

\makeatletter
\def\ps@IEEEtitlepagestyle{%
	\def\@oddfoot{\mycopyrightnotice}%
	\def\@evenfoot{}%
}
\def\mycopyrightnotice{%
	\begin{minipage}{\textwidth}
		\centering\tiny{\copyright{} 2021 IEEE.  Personal use of this material is permitted.  Permission from IEEE must be obtained for all other uses, in any current or future media, including reprinting/republishing this material for advertising or promotional purposes, creating new collective works, for resale or redistribution to servers or lists, or reuse of any copyrighted component of this work in other works.}
	\end{minipage}
	\gdef\mycopyrightnotice{}
}

\begin{document}

\title{A Self-Configurable Grouping Method for Integrated Wi-SUN FAN and TSCH-based Networks
}

\author{
    \IEEEauthorblockN{
    Xinyu Ni\IEEEauthorrefmark{1}, 
    Michael Baddeley\IEEEauthorrefmark{1},
    Nan Jiang\IEEEauthorrefmark{1},
    Yichao Jin\IEEEauthorrefmark{1}}
    
    \IEEEauthorblockA{\IEEEauthorrefmark{1}Toshiba Europe Ltd., Bristol, United Kingdom, \\ Email:\{Xinyu.Ni, Michael.Baddeley, Nan.Jiang, Yichao.Jin\}@toshiba-bril.com
    }
    
\vspace*{-0.6cm}
}
\maketitle


\begin{abstract}
Recent applications in large-scale wireless mesh networks (WSN), e.g., Advanced Metering Infrastructure (AMI) scenarios, expect to support an extended number of nodes with higher throughput, which cannot be sufficiently supported by the current WSN protocols. Two prior protocols, Wi-SUN Field Area Network (Wi-SUN FAN) and IETF 6TiSCH standards, are popularly used that are respectively based on asynchronous Carrier Sense Multiple Access / Collision Avoidance (CSMA/CA) and IEEE 802.15.4 Time Scheduled Channel Hopping (TSCH). However, the former one with CSMA/CA can be prone to the Hidden Node Problem (HNP) that leads to a degradation of reliability, while the latter one with TSCH avoids HNP by using synchronously scheduling but requires massive control signalling that degrades the upper-bound of throughput. Accordingly, this paper tackles the challenge of how to improve the upper-bound of throughput without loss of reliability. To do so, we first present an in-depth evaluation of reliability and throughput between the two existing standards via system-level simulation. We then propose a self-configurable grouping (SCG) method to cluster nodes without using location information of each node. This SCG method eliminates around 99\% HNP in CSMA/CA, thus greatly improves its reliability with maintaining the relatively high upper-bound of throughput. Our results show that the SCG method almost doubles the network throughput compared with both 6TiSCH and Wi-SUN FAN in heavy traffic scenarios, while providing extremely high reliability of more than 99.999\%.
\end{abstract}

\vspace{0.2cm}

\section{Introduction}
\label{sec_intro}

Wireless mesh networks (WSN) are expected to offer connectivity for massive number of nodes with high reliability. However, recent applications, e.g., Advanced Metering Infrastructure (AMI) scenarios, expect to support nodes with higher throughput, which cannot be sufficiently supported by the current WSN protocols. One of the popular protocols is the Field Area Network (FAN) profile from the Wireless Smart Utility Network (Wi-SUN) Alliance, which is based on asynchronous Carrier Sense Multiple Access / Collision Avoidance (CSMA/CA) \cite{std_wisun_fan}. The reliance on CSMA/CA introduces a general problem - Hidden Node Problem (HNP), which can lead to a degradation of reliability. In contrast, based on IEEE 802.15.4 Time Scheduled Channel Hopping (TSCH), another popular protocol produced by Internet Engineering Task Force (IETF), namely, IPv6 over TSCH (6TiSCH) achieves relatively higher reliability by using scheduling \cite{6tisch_ietf_architecture}. However, its upper-bound throughput is greatly degraded due to the reason that the massive channel is occupied by control signalling for synchronously scheduling. In this paper, we target addressing the problem of hidden nodes in CSMA/CA-based networks by using TSCH scheduling so as to improve the upper-bound throughput of WSN without reducing its reliability.

The Wi-SUN FAN protocol is widely adopted as the key standard for recent Advanced Metering Infrastructure (AMI) providers~\cite{wisun-overview}. Based on low-power IEEE 802.15.4-2015~\cite{ieee_std_802154_2015}, the current Wi-SUN specification employs CSMA/CA, where each node contends for channel access. However, this approach can incur a high probability of packet losses due to the well-known HNP~\cite{rahman2006hidden}. In details, this problem happens when two nodes attempting concurrently transmissions to the same receiver. If both nodes were hidden from each other, Clear Channel Assessment (CCA) check performed on each node targeting on channel occupation detection would fail. Their packets subsequently collide at the receiver, where packets from both are unable to be decoded correctly. HNP can therefore cause performance degradation to both network throughput and Packet Delivery Ratio (PDR). For instance, the large-scale WSN is generally limited with up to 2000 nodes per data concentrator in AMI applications \cite{tepco_landys_2017}. As increasing the scale of the network is required by modern WSN applications, tackling the challenge of HNP for asynchronous CSMA/CA protocols becomes even more urgent.

In contrast, 6TiSCH makes use of advanced scheduling algorithms to coordinate each communication in a non-conflicting manor, which enables communication in a deterministic manner. The current 6TiSCH standard generally recommends a fixed slot duration in order to provide synchronization guard times and complete bi-directional communication between the sender and the receiver. The duration of a slot is selected according to the requirements of traffic and the computational capability of the hardware. In general, serving a packet with 6TiSCH needs to consume double of channel resource than that with the Wi-SUN's CSMA/CA approach without considering HNP \cite{ieee_std_802154_2015,harada2017ieee}. Therefore, the upper-bound throughput of 6TiSCH is limited by its high requirement of control signalling.

In this work, we target on addressing the problem of hidden nodes in CSMA/CA-based networks by using TSCH scheduling. We present an in-depth performance evaluation for both Wi-SUN FUN and 6TiSCH techniques based on system-level simulation. The simulation showcases that the performance of CSMA/CA is degraded due to suffering from collision, while the upper-bound performance of TSCH scheduling is limited due to its heavy control signalling. We argue that, according to results of evaluation, one may use TSCH scheduling to cluster discoverable CSMA nodes within small groups to avoid nodes hidden from each other, thus can improve overall throughput by solving HNP. To do so, we propose a self-configurable grouping (SCG) method to cluster nodes without using location information of each node. By using the proposed method, 99$\%$ HNP has been avoided that reduces the most collision within the network, thus achieves around double throughput than solely using 6TiSCH approaches.

The rest of the paper is organized as follows. Section \ref{sec_background} introduces Wi-SUN FUN and 6TiSCH network as well as compares their performance by system-level simulation. Section \ref{sec_approach_grouping_method} proposes SCG method in detail to address the problem of hidden nodes and provides simulation results. Finally, section IV concludes the paper.

\section{Scalability Evaluation of Wi-SUN and 6TiSCH}
\label{sec_background}

In this section, we first briefly introduce the protocols of physical (PHY) and medium access control (MAC) layers in asynchronous Wi-SUN and synchronous 6TiSCH. After that, we compare their reliability and scalability by evaluating the throughput and the PDR for different traffic scenarios under system-level simulations. 

\vspace*{-0.2cm}
\subsection{Asynchronous Wi-SUN}
The lower-layer specification of Wi-SUN wireless networks are defined by CSMA/CA according to IEEE 802.15.4 technical standard, which specifies PHY/MAC layers for low-rate wireless personal area networks (LR-WPANs) \cite{ieee_std_802154_2015}. An unslotted CSMA/CA principle is used, where nodes can access to the network once backlogged without any scheduling procedure in an asynchronous manner \cite{wisun_paper2019}. Due the the unscheduled principle, collision may occurs, i.e., two or more nodes select the same channel at the same time, that leads to a transmission failure. To avoid the currency of collision, a \textit{back-off} mechanism is introduced that postpones the access attempts of nodes after detecting a busy channel, which can guarantee the transmission of the current one.

In details, If either a busy channel was detected by a node or collision was occurred, the node would postpone its transmission for a duration (measured by slots) that is uniformly selected in random from a range $[c_\text{NB}+c_\text{BE}, c_\text{NB}+c_\text{BE}+{2^{c_\text{CW}}}-1]$. Here, $c_\text{NB}$ is a constant, $c_\text{BE}$ is a back-off exponent selected by the root node within the range $[1, 3]$ with a default value 3, and $c_\text{CW}$ is the number of transmission attempts \cite{ieee_std_802154_2015}. At the end of the back-off duration, the node performs a single successful CCA to check whether the channel is busy by detecting energy above certain energy threshold or spreading signal generated by neighboring nodes. Since each node has a fixed range of communication area, CCA check cannot cover the whole network. Therefore, neighboring nodes out of the communication range cannot be detected, resulting in the well-known HNP. Additionally, if two nodes perform CCA check at the same time, there is opportunity that the CCA may report an idle medium, which can also lead to packet collision.

\subsection{6TiSCH Synchronous MAC Layer}

In contrast, 6TiSCH provides scheduling services based on the synchronous IEEE 802.15.4-2015 TSCH~\cite{ieee_std_802154_2015}. A 6TiSCH scheduler orchestrates communications over individual links in an optimized and non-conflicting manner. In the time domain, the schedule operates in a Time Division Multiple Access (TDMA) manner, whereas in the frequency domain it divides the wireless spectrum into multiple channels. A schedule is formed by assigning \textit{time slot} and channel offsets to each communication link, and specifying which node should transmit or receive data to/from its scheduled counterpart. However, the standard never specified \textit{how} these resources are scheduled, which has consequently been the focus of the IETF 6TiSCH Working Group \cite{6tisch_ietf_architecture}.

In 6TiSCH, all the nodes are synchronized and can only transmit and receive packets in scheduled equal-long time slots to avoid collision and maintain reliability \cite{6tisch_2014}. The length of the time slot is typically selected from 10 to 15 ms according to the requirements of application \cite{ieee_std_802154_2015}, while its maximal length is limited by the the time of transmitting the largest MAC frame (127 bytes long with overhead 25 bytes \cite{montenegro2007transmission}) plus the time of the following acknowledge (ACK) frame for informing successful transmission. For instance, if the nodes operate in the 2.4GHz frequency band, it takes approximately 4 ms to transmit the maximum size MAC frame and extra 1 ms for the ACK. If the time slot is set to 10 ms, then 5 ms is used for radio turnaround, packet processing, and security operations \cite{rfc7554}.

\subsection{Simulation Results and Evaluation}
\label{sec_approach_analysis}

\begin{figure*}[ht]
    \centering
    \includegraphics[width=.9\textwidth]{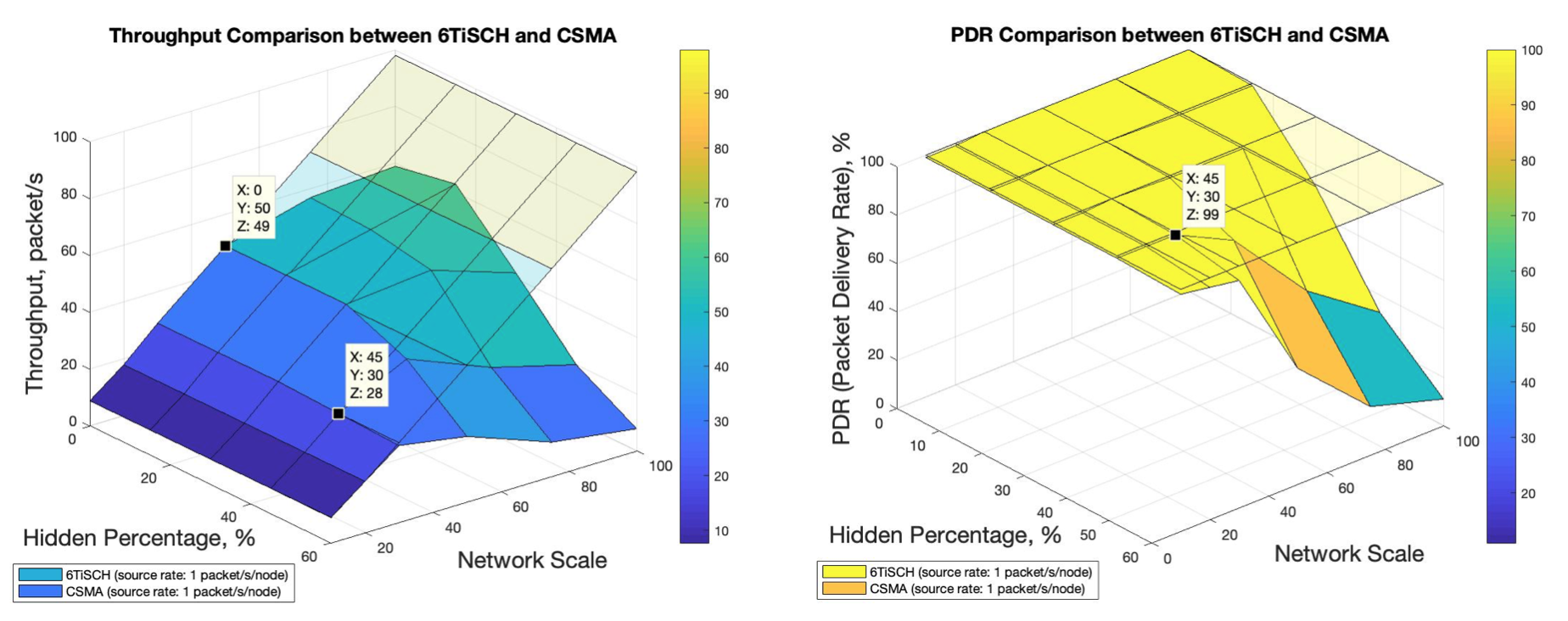}
    \caption{Throughput (left) and PDR (right) of 6TiSCH and CSMA in a 100 node scenario where traffic source rate is fixed with each node generating 1 per second. CSMA throughput increases linearly as the transparent plane, while 6TiSCH drops as the network scales. 6TiSCH maintains 100\% PDR as the transparent plane, while CSMA drops as the network scales.}
    \vspace{-0.2cm}
    \label{fig:03_tput_hnp_scale_comparison}
\end{figure*}

\begin{figure*}[ht]
    \centering
    \includegraphics[width=.9\textwidth]{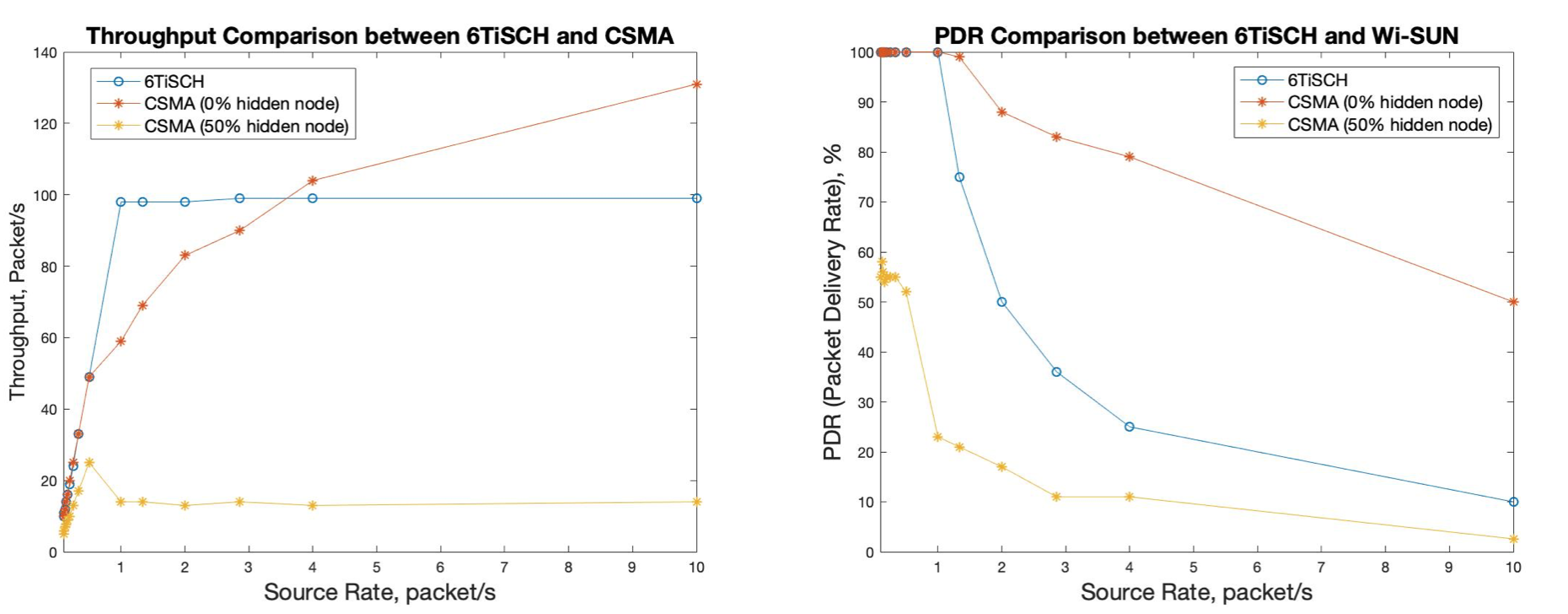}
    \caption{Throughput (left) and PDR (right) of 6TiSCH and CSMA in a 100 node scenario over different source rates. CSMA HNP is compared at 0\% and 50\%, while in 6TiSCH all nodes are scheduled to not contend.}
    \vspace{-0.4cm}
    \label{fig:03_tput_6tisch_csma}
\end{figure*}

In this subsection, system-level simulations are conducted using Cooja, i.e., a simulation tool for the Contiki operating system (OS)~\cite{contiki}, to evaluate the performance of Wi-SUN and 6TiSCH networks. A single-hop simulation scenario is considered, where each node is configured to send packets to a coordinator in a unicast communication. As we targeting on a MAC level evaluation, the capture effect of physical channels, e.g., noise and multipath fading, are omitted by considering a receive probability with 100\%. For clarity, a hidden node percentage for the whole network is referred, which is calculated by averaging the sum of hidden node percentage between each communication link, which is represented by 
\begin{equation}\label{eq1}
P_n = \frac{1}{n} \sum_{i,j=1}^{i=I,j=J}P_{ij}.
\end{equation}
In \eqref{eq1}, $n$ is the total number of links in the network, and $P_{ij}$ is the hence hidden node percentage for a link of uni-cast transmission from node $i$ to node $j$. This probability can be obtained by
\begin{equation}
P_{ij} = \frac{N_i - N_i \cap N_j}{N_i}, 
\end{equation}
where $N_i$ and $N_j$ are the number of neighbors of node $i$ and $j$, respectively. 

We consider a scenario for both 6TiSCH and CSMA networks with an arbitrary number of nodes randomly located in the area of a plane. The location of each node is fixed along time. We consider a periodic traffic, where every node generates an inter-arrival packet per second waiting to be served. The time is divided into slots with a fixed duration of 10 ms, where each slots contains $F$ number of independent channels. Note that the scheduling and control signalling for a packet are conducted in a single slot. Therefore, the maximal throughput of this network is 100 packets/second. For CSMA/CA, the back-off parameters $c_\text{NB}$ and $c_\text{BE}$ are set to 5 and 7, respectively.

Figure \ref{fig:03_tput_hnp_scale_comparison} illustrates the throughput (left) and PDR (right) of 6TiSCH and CSMA networks versus hidden percentage and network scale with a fixed source rate (traffic arrival rate) of each node that generates 1 packet per second. We can observe that the throughput of 6TiSCH is proportional to the product of source rate and network size without being affected by the hidden node problem in difference network scales. This performance occurs due to the fact that the centralized scheduling algorithm is implemented, where the coordinator has the full knowledge of the whole network. By doing so, it can perfectly schedule the transmission of each node with a deterministic time slot for guaranteed packet transmission, thus avoids collisions. Wi-SUN, however, uses contention based CSMA/CA for data transmission. We observe that, when the hidden percentage is small, the throughput is slightly degraded while PDR remains at a high level due to that the re-transmission times is high enough. On the other hand, if the hidden percentage is large, even in a moderate network scale, both throughput and PDR will drop dramatically.

Figure \ref{fig:03_tput_6tisch_csma} illustrates the throughput (left) and PDR (right) of 6TiSCH and CSMA networks versus different source rates ranged from 1to 10 packet/1s/node in a scenario with 100 nodes. As can be seen, 6TiSCH can always maintain a throughput close to its upper bound (i.e. approximately 98 packets/s in 100-node scenario without including time slots are used for synchronization) and the linear relationship between PDR and source rate indicates the 100\% PDR for packet transmissions which are scheduled properly. Due to HNP, Wi-SUN with high hidden percentage suffers from lower throughput than 6TiSCH, which indicates the performance degradation of collision problem. 

However, we can also observe that the throughput of Wi-SUN with 0 hidden percentage achive a higher throughput than 6TiSCH when the source rate larger than 4. This is due to the reason that the CSMA/CA contributes to a more simplified protocol than the scheduling method in 6TiSCH requiring less control signalling, which leads to a higher channel occupancy of data signals. For instance, CSMA/CA based on IEEE 802.15.4 with OQPSK-DSSS can achieve 256 kbps that only requires 4 ms channel duration, which maintains a similar throughput in 6TiSCH with 10 ms channel duration \cite{ieee_std_802154_2015}. Therefore, we argue that if the CSMA was immune from the severe effect of HNP, the network would achieve higher throughput in heavy traffic scenario than 6TiSCH. In the next section, we propose a heuristic solution to solve HNP without increasing the complexity of the network.

\section{Self-Configurable Grouping Method}
\label{sec_approach_grouping_method}

Large scale mesh networks can be distributed in a wide area. This can result varying performance within different subsections of the network, depending on the local environment as well as application requirements. The density of the network may change, thus subsections may experience localized interference, and suffer low data rates. For the subsections with bursty traffic, a contention-based MAC approach can provide higher upper bound of throughput due to its lower requirements of control signalling, which might be better serve the bursty data.
For the subsections with lighter traffic, one could use a TSCH approach to save power and improve reliability by scheduling non-contending transmissions. We therefore propose a method to group different subsections of the network in terms of different throughput requirements, guaranteeing both throughput and reliability requirements.  

\captionsetup{singlelinecheck=false} 
\begin{figure}[t]
\vspace*{-0.0cm}
    \begin{center}
        \begin{minipage}[t]{0.49\textwidth}
    \centering
        \includegraphics[width=1\textwidth]{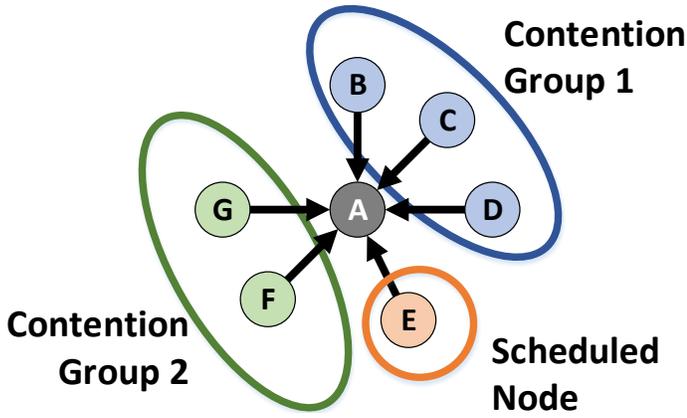}
        \vspace*{-0.0cm}
        \caption{Grouping algorithm example. Neighboring nodes are clustered into contending groups.}
                \label{fig:GE1}
        \end{minipage}
    \end{center}
    \vspace*{-0.3cm}
\end{figure}

\subsection{SCG Algorithm}
We consider a multi-hop mesh network composed of a coordinator node located at the center of an plane area $R^2$, and a set of static nodes randomly located in this area. Considering the nodes can support both CSMA/CA and scheduling protocol through TSCH. Both techniques share $R$ number of channels. Devices can access to the network using either contention-based CSMA/CA or scheduling-based TSCH.

To reduce hidden nodes probability, neighbouring nodes with full detection among each other can be marked as a group to share the same channel, which can help to avoid HNP. As illustrated in Figure \ref{fig:GE1}, two contention-based groups are created, as well as a single node that is scheduled normally through TSCH. Neighboring nodes (e.g. Group 1) are grouped and scheduled to avoid hidden nodes common to the group (e.g. Group  2). This allows the group to take advantage of shorter CSMA transmission periods, while avoiding the reliability hit from HNP.

\begin{algorithm}[t]  
    \footnotesize
    \caption{Grouping Algorithm} 
    \label{alg:grouping_algorithm}
    \noindent\hspace*{\algorithmicindent} \textbf{Input:} Node struct an information tensor $N$ containing parent and child node info $N.childNodes$, traffic $N.traffic$, and neighbours $N.neighbours$;
    \begin{algorithmic}[1]
        \For {Node $i= 1 \to I$}  
            \State Node $i$ broadcast neighbouring calculation requests;
            \State Node $i$ receives ACK from neighbours with their ID;
            \State Node $i$ formulates a neighbour's ID list $l_i$ according to receptions;
            \State Node $i$ transmits list $l_i$ to the coordinator;
        \EndFor
        \State Coordinator construct grouping matrix $CNI = []$, and ID list contains all the IDs of the nodes if they have child nodes, $L_ID = [l_1, l_2, \cdots, l_I]$;
            \For {each $i\in L_ID$}     
                \State $grouping\_done = 0$;
                \State $CNI(1, :) \leftarrow N(i).childNodes$;
                \State $CNI(2, :) \leftarrow  N(CNI(1,:)).neighbours$;
                \State Sort CNI in ascending order;
                \While {$grouping\_done \neq 1$}
                    \State $NodeS = CNI(1,1)$;
                    \State $group\_temp = N(NodeS).Neighbour$;
                    \State $filtering\_done = 0$;
                    \While{$filtering\_done \neq 1$}
                        \State $CI(1,:) = group\_temp$;
                        \State $CI(2,:) = HNPCal(CI(1,:))$;
                        \State Sort CI in descending order;
                        \If {$CI(1,1)==0$ AND $CI \neq []$}
                            \State $filtering\_done = 1$;
                        \EndIf
                    \EndWhile
                    \State $group\_temp = CI(1,:)$;
                    \If {$N(group\_temp).layerID == 2$ OR $N(group\_temp).layerID == 3$}
                        \If{$size(group\_temp) > 2$}
                            \State $group\_temp = group\_temp(1:2)$;
                        \EndIf
                    \Else
                        \State $group\_temp = group\_temp(1:4)$;
                    \EndIf
                \EndWhile
            \EndFor
            \State Coordinator publish $group\_temp$ to each node.
    \end{algorithmic}  
\end{algorithm}

\begin{figure*}[ht]
\setcounter{figure}{4}
    \centering
    \includegraphics[width=0.9\textwidth]{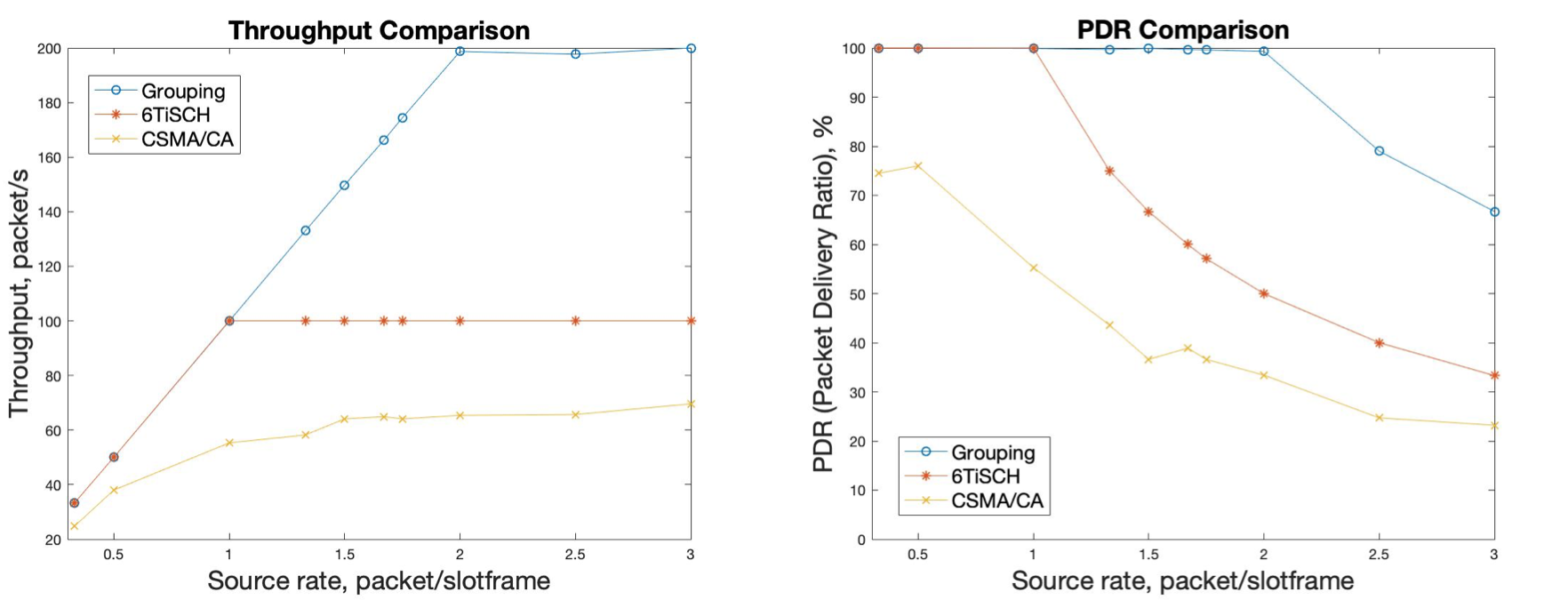}
    \caption{a) Throughput comparison for Grouping, 6TiSCH and CSMA in 300-node network at different source rate (left). b) PDR comparison for Grouping, 6TiSCH and CSMA in 300-node network at different source rate (right).}
    \label{fig:04_tput_pdr_comparison}
\end{figure*}

However, the location information of each node is hardly obtained by themselves as well as the coordinator. We consider the location of each node is unknown by themselves, an SCG approach is proposed then to group nodes aiming at improving overall throughput of the whole network by reducing hidden node probability at most. At the deployment stage, the coordinator and nodes execute SCG approach by communication among each other using conventional TSCH scheduling method. A initial flag is multi-cast by coordinator to all nodes to initialize SCG calculation. The algorithm is detailed in Algorithm~\ref{alg:grouping_algorithm}, and its implementation details are given as follows:

\begin{itemize}
    \item Step 1: All nodes progressively multi-cast a signal with ACK feedback to calculate their potential neighbours, calculate a neighbour list, and forward the list to the coordinator. The child node with least number of neighbors is chosen as the start point. Hence, all the nodes which are neighbors of that node, and have the same parent, can be added in a temporary group list.
    \item Step 2: the HNP of each node in the temporary group list is calculated and sorted in descending order. 
    \item Step 3: the node with highest HNP is removed from the temporary group list and HNP of all the nodes left in the list is calculated again. 
    \item Step 4: Step 2 and 3 are repeated until all the nodes in the temporary group list are not hidden from each other or the temporary group list is empty.
    \item Step 5: If the temporary group size exceeds the limit, the most distant neighbors will be removed from the temporary group list. 
\end{itemize}

\captionsetup{singlelinecheck=false} 
\begin{figure}[t]
\vspace*{-0.0cm}
    \begin{center}
        \begin{minipage}[t]{0.48\textwidth}
    \centering
        \includegraphics[width=1\textwidth]{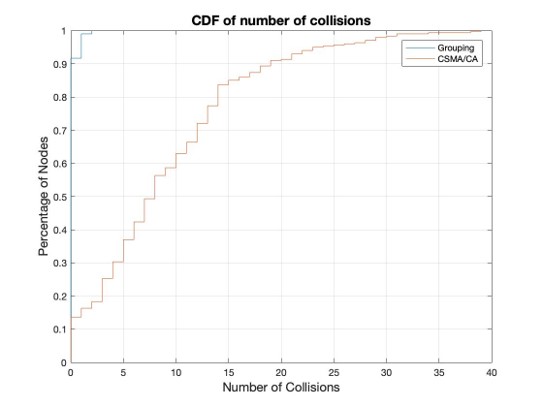}
        \vspace*{-0.5cm}
        \caption{Comparing the CDFs of the number of collisions between conventional CSMA/CA and the proposed SCG algorithm in a scenario with 300 nodes randomly located in a square area with a side length of 100 meters.}
                \label{fig:CDF}
        \end{minipage}
    \end{center}
    \vspace*{-0.8cm}
\end{figure}

Fig. \ref{fig:CDF} shows the cumulative distribution function (CDFs) of the number of collisions between conventional CSMA/CA and the proposed SCG algorithm in a scenario with 300 nodes randomly located in a square area with a side length of 100 meters. We can observe that around 99$\%$ collisions are eliminated by using the proposed SCG algorithm. 



\begin{figure*}[ht]
\setcounter{figure}{5}
    \centering
    \includegraphics[width=0.95\textwidth]{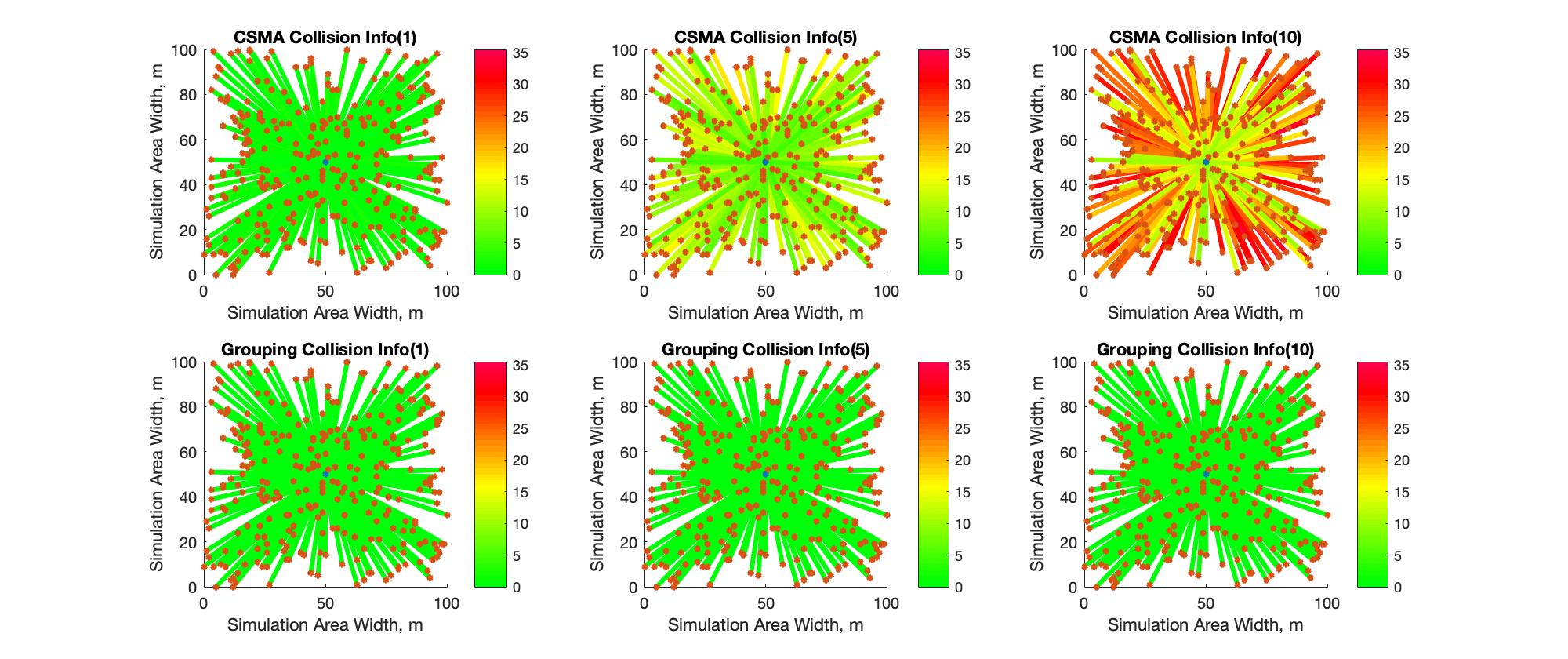}
    \caption{Collision Comparison between CSMA/CA and Grouping at different source rates in a 300-node scenario}
    \label{fig:04_grouping_v_csma}
\end{figure*}

\label{sec_results}
\subsection{Numerical Results and Evaluation}
In this subsection, a simulation of large-scale multi-hop network is conducted to evaluate the performance of the proposed SCG algorithm using Matlab. The scheduler used for group communication~\cite{AMUS} is compared with a known 6TiSCH scheduler~\cite{accettura2015decentralized}, as well as unslotted CSMA/CA. We consider a multi-hop mesh network with 300 nodes randomly located in a square area with a side length of 100 meters. Packets are periodically generated by each node, and the packet size is fixed to 50 bytes. In CSMA/CA scenario, the IEEE 802.15.4 with OQPSK-DSSS is used, which can achieve 256 kbps, and channel duration is set to 4 ms that is sufficient to serve the generated packets. For TSCH, a standard 10 ms channel duration is set according to \cite{ieee_std_802154_2015}. Note that the communication is assumed to be bi-directional and HNPs of both uplink and downlink communication are taken into consideration.

Figure \ref{fig:04_tput_pdr_comparison} illustrates both the throughput and PDR comparison among the three methods. The hidden node probability is about 43\%. When the source rate is low, both grouping and 6TiSCH can maintain high PDR while CSMA/CA has only approximately 75\% reliability due to severe collisions. The throughput of both grouping and 6TiSCH methods increase with the increase of source rate until they meet the upper bound. Compared with the 6TiSCH method, the grouping method can reach higher upper bound due to the reason that it requires less control signalling. It can also be observed that, due to the scheduling strategy, the 6TiSCH method can remain the highest throughput even the source rate is keeping increase. Accordingly, the grouping method also shows the similar trend as the 6TiSCH method, which sheds light on that the HNP is closely eliminated by the proposed SCG algorithm.

The number of collisions of each node in one simulation episode is shown in Figure \ref{fig:04_grouping_v_csma}. It can be observed that, when the source rate is low, both CSMA/CA and grouping methods have rare collisions, while, when the source rate is high, CSMA/CA method suffers from a significant collision situation.

\section{Conclusion}
\label{sec_conclusion}
In this paper, we compared CSMA/CA and 6TiSCH via system-level simulation and developed an SCG algorithm to solve HNP in CSMA/CA networks. We evaluated that the HNP limits the throughput and PDR of CSMA/CA due to collisions, while the complexity of control signalling in 6TiSCH limits its upper bound of throughput. The proposed SCG algorithm can remarkably solve the problem of HNP that can approximately double the throughput than that of 6TiSCH without reducing its reliability. This is because, with the help of SCG algorithm, neighbouring nodes are grouped with no HNP inside each group, thus can delivery packets in an unslotted CSMA/CA manner without degradation of HNP. The future work will study the latency limitation of the grouping method CSMA/CA based mesh networks. 



\bibliographystyle{IEEEtran}
\bibliography{references}
\end{document}